# Quantum mechanics vs relativity: an experimental test of the structure of spacetime

**S. A. Emelyanov**

*Ioffe Institute, St. Petersburg 194021, Russia*

*E-mail: sergey.emelyanov@mail.ioffe.ru*

**We have performed an experimental test under the conditions of which quantum mechanics predicts a spatially-discontinuous single-particle transport. The transport is beyond the relativistic paradigm of movement in Cartesian space and therefore may well be nonlocal. Our test has demonstrated that such transport does exist. This fact opens the door for realistic interpretation of quantum mechanics insofar as the requirement of Lorentz invariance appears inapplicable to any version of quantum theory. Moreover, as quantum mechanics proposes a particle dynamics beyond the relativity, it automatically requires an adequate "quantum" concept of spacetime, for which the relativistic concept is only a limiting case. The quantum concept allows absolute simultaneity and hence revives the notion of absolute time. It also goes beyond the relativistic curvilinear Cartesian order of space to account for quantum phenomena such as the discontinuity and the nonlocality in the spirit of Bohm's concept of the implicate order.**

PACS numbers: 03.65.-w, 03.67.-a, 04.20.Gz

## 1. Introduction

Quantum phenomena cannot be fully described in terms of conventional Cartesian space but only in terms of a multidimensional configuration space. This fact was realized by the very founders of quantum mechanics and discussed in the historical fifth (1927) Solvay Conference [1]. At the Conference, both Lorentz and Einstein repeatedly urged participants to use the framework of three-dimensions in their theoretical approaches to quantum mechanics. However, Schrödinger with his wave mechanics as well as de Broglie with his pilot-wave theory was unable to follow this appeal.

The general concern about the framework of quantum mechanics is quite evident. If the framework of Cartesian space appears too narrow to describe quantum phenomena, than the relativistic concept of spacetime may be at stake. Nevertheless, most participants, such as Bohr, Heisenberg, Dirac and Pauli were convinced in the fundamental impossibility to describe, for instance, quantum interference in terms of Cartesian space. However, they proposed an original way to reconcile relativity with quantum mechanics, that is, to sacrifice realism in quantum mechanics. Since that time, their unrealistic (Copenhagen) interpretation of quantum mechanics is known as just the quantum mechanics [2-5]. Following to this interpretation, we never can talk meaningfully about quantum object in itself but only in the context of a specific experiment. As was stressed by Bohr, any attempts to treat quantum object separately will lead to misconceptions [2].

However, Bohr's approach has at least two dark points emphasized, for example, by Einstein [6-8]. The first one is the role of observer who, for a certain time, mysteriously pulls quantum object out of nothing and, then, no less mysteriously the object goes back to nothing. Today it is known as the measurement problem which is generally resolved in such a way that mysticism is regarded as inevitable attribute of quantum world despite Einstein's admonition that such an approach contradicts the very spirit of science as an objective knowledge about the reality "out there". Following to Einstein, quantum mechanics appears thus not strictly a scientific theory but rather a mathematical tool to predict experimental observations. At present, the common way to avoid this problem is the so-called "shut-up-and-calculate" approach though it hardly looks like a fully scientific one.

The second dark point is the problem of a transition from classic to quantum description. It is not the problem of a criterion but rather a more general problem of how to understand such transition if, according to Bohr's concept, it is actually the transition from realism to mysticism. It is indeed hard to believe that a purely physical criterion, regardless of its specific content, could provide a "switching" between mutually-exclusive methods. On the other hand, if we suppose that classic description is a limiting case of quantum description without any "switching", than we have come to paradoxes. The first is the well-known paradox of Schrödinger's cat. The other one was expressed by Einstein in a just one sarcastic phrase: "Do you really think the Moon isn't there if you aren't looking at it?"

However, despite all objections, further course of events seems to confirm Bohr's concept. First of all it is the observation of nonlocal correlations under the conditions of Einstein-Podolsky-Rosen *gedanken* experiment that was originally conceived just to demonstrate an incompleteness of quantum theory [9]. Actually, in the light of Bell's theorem, EPR experiments play the role of a test that rules out any local hidden-variable theories and even the nonlocal ones covered by so-called Leggett theorem [10-13]. Moreover, even if a realistic version of quantum theory, most notably de-Broglie-Bohm theory, does not contradict EPR experiments, it nevertheless faces a truly formidable difficulty with nonlocality which, in a combination with realism, seems incompatible with relativity [14-20]. Today this incompatibility is expressed through the so-called quantum dilemma that roughly sounds as "either nonlocality or realism" [21-23]. Thus, the potential conflict with Einstein's relativity ironically appears just the stumbling-stone for his dream of a realistic quantum theory [24-25].

Along with the violation of realism, the key point that allows one to maintain a fragile balance between quantum mechanics and relativity is the impossibility of a nonlocal signaling through EPR-type nonlocal correlations. It is the subject of so-called no-communication theorem that allows a purely mathematical proof [26-28]. So, as the vast majority of articles devoted to quantum nonlocality ultimately address EPR correlations, it is generally believed that the impossibility of nonlocal signaling is unavoidable and moreover is a deeply fundamental thing. Perhaps only John Bell (in his celebrated work [16]) found "… disturbing … the impossibility of "messages" faster than light, which follows from ordinary relativistic quantum mechanics in so far as it is unambiguous and adequate for procedures we can actually perform…" though this phrase is still regarded rather as a quirk and no one take it seriously.

## 2. Implicit nonlocality of single-particle wavefunction: a *gedanken* experiment

However, an alternative type of quantum nonlocality nevertheless exists though it is much less common in literature. In passing, it was mentioned just in the above-mentioned Bell's work. We mean implicit nonlocality of a wavefunction itself. The possible reason for a much less attention to this type of nonlocality is that it has never been associated with a reasonable experiment. But it does not mean that such experiment is impossible even in principle. As an example, consider a *gedanken* experiment leaving yet aside the problem of its realizability.

Let there is a quantum system, the eigenstate of which is a macroscopic-scale one-dimensional circular orbit *C* (Fig. 1). In the system, we can easily select two spatial domains so that both are crossed by the orbit *C* but remote from each other by a macroscopic distance *L*. Let the first (Alice's) domain contains also a local level *A* while the second (Bob's) domain contains a local level *B* together with a number of local scatterers capable to provide transition $C \to B$ during characteristic time ($\tau_{C \to B}$).

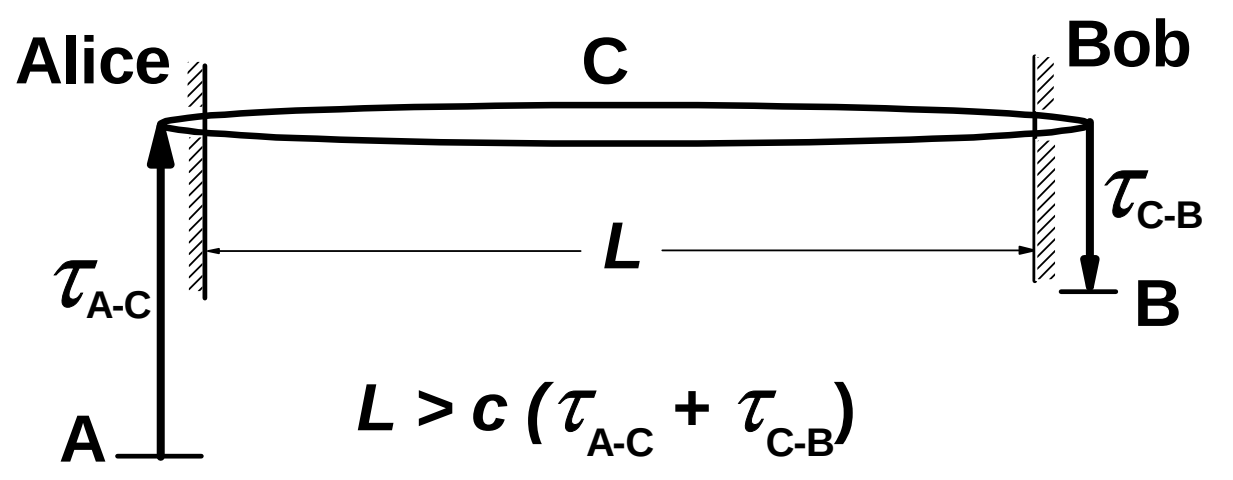

**Fig. 1.** ***Gedanken*** **experiment with a nonlocal transport due to implicit nonlocality of single-particle wavefunction.** ***C*** **– macroscopic-scale circular quantum orbit,** ***L*** **– the distance between Alice and Bob,** ***A*** **and** ***B*** **– local quantum levels belonged to Alice's and Bob's domain, respectively. Vertical arrows denote local quantum transitions.**

Initially, only the lowest level (*A*) is occupied by a single electron. Then, let Alice exposes her domain to light capable to provide $A \rightarrow C$ transition during characteristic time $\tau_{A\rightarrow C}$. In this case, the electron should appear in the level *B* during the time ($\tau_{A\rightarrow C} + \tau_{C\rightarrow B}$) and actually it is nothing but a transport from *A* to *B*. In accordance with the principle of quantum indeterminacy, the distinct feature of the transport is that the electron appears in the level *B* *without* passing through intermediate positions between *A* and *B*. In other words, we face a *spatially-discontinuous* transport which is beyond the very paradigm of movement in Cartesian space. Its characteristic time ($\tau_{A\rightarrow C} + \tau_{C\rightarrow B}$) clearly has nothing to do with the time spent in overcoming the distance *L* and thereby is independent of *L*. This means we can easily provide the conditions under which $L/\tau_{A\rightarrow C} + \tau_{C\rightarrow B}$ is higher than the speed of light and it would be not in a contradiction with relativity simply because we are beyond the very paradigm of movement in Cartesian space. We are also beyond the no-communication theorem relevant only to quantum entanglement. Moreover, it is clearly seen that the lengthscale of spatial discontinuity is determined by that of the orbit $C$ and generally speaking it has not any limitations of a fundamental character.

## 3. The idea of realization

Consider now the problem of realizability of our *gedanken* experiment. In the Bohr-Einstein times, one would indeed think of its fundamental non-realizability. Nowadays, however, one would provide a counter-example which moreover could be realizable right now. Indeed, consider a macroscopic quantum system known as the integer quantum Hall (IQH) system [29-30]. According to generally-recognized Laughlin-Halperin's theory, each IQH system contains the so-called current-carrying states extended along the perimeter *regardless* of system size [31-32]. In these states, electrons behave as spontaneous quasi-one-dimensional currents with cross-section of the order of microscopic cyclotron radius. Roughly, the states are due to the crossing of external quantizing magnetic field and in-plane electric field that always occurs in the vicinity of sample edges. In fact, the states may well be regarded as quantum orbits which seem suitable to perform the desired experiment. However, the problem is that in optical experiments, such as that one in Fig. 1, it is hard to distinguish the effect of a small number of current-carrying states from the effect of a gigantic number of localized states in the system interior. This fact is known, for instance, from photo-conductivity measurements performed in the IQH regime [33].

However, the situation can be improved in two ways. The first one is that we can try to provide Laughlin-Halperin-type states not only close to system edges but also in the interior. Indeed, if IQH system has an asymmetric confining potential, then there should be the so-called built-in electric field which is an analogue of the in-plane field near the system edges. In this case, if the external magnetic field has not only quantizing but also an in-plane component, then the crossing of electric and magnetic field will occur in the system interior as well. This could be the reason for the Laughlin-Halperin-type states there.

Indirectly, our guess is supported by theoretical calculations with a simplified model of infinite IQH-like system [34]. According to these calculations, if there is a gradient of electric potential in the *Z*-direction while external magnetic field has both quantizing (*Z*) and in-plane (*X*) components, then electrons' in-plane velocities may be nonzero in the *Y*-direction. In this case, energy spectrum consists of a series of so-called Landau subbands shifted in *k*-space ($\varepsilon(k_y) \neq \varepsilon(-k_y)$) so that their shift depends on the Landau quantum number as well as on the system toroidal moment which is a cross product of electric and magnetic fields [35-36]. In the *X*-direction, however, electron's wavefunctions are strongly restricted by their cyclotron radius ($r$). Moreover, their *X*-coordinates are in a correlation with their wave vector ($x_0 = -k_y r^2$), just like in the case of conventional IQH system. It follows that the electrons are spatially-separated one-dimensional spontaneous currents flowing in opposite directions along *Y*-axis and they are strongly reminiscent the Laughlin-Halperin's edging currents. The only problem is that the model supposes an endless IQH system and it is not yet clear whether or not the currents exist in real system of a finite size. It is thus still our guess that the currents do exist in such a way that they are closed through the true Laughlin-Halperin's edging states (Fig. 2A). In the case our guess will prove to be valid, we do obtain a number of spatially-separated one-dimensional macroscopic orbits.

The second way to improve situation is that we could try to find a type of optical measurements, which are sensitive only to current-carrying states but insensitive to conventional (localized) IQH states. In other words, we should observe an output signal only if the energy spectrum of localized states (Fig. 2B) truly transforms into the energy spectrum of Laughlin-Halperin-type states (Fig. 2C). Fortunately, the desired method of optical measurements does exist. To describe it, consider the scheme of optical transitions between the Landau subbands. In conventional unbiased IQH system with localized states, these transitions cannot result in an in-plane current because of the in-plane velocities are zero (Fig. 2B). By contrast, in Fig. 2C their in-plane velocities are nonzero and moreover they are not exactly the same at initial and final state of an optical transition. Therefore, a nonzero net current could potentially emerge even in unbiased system under the cyclotron resonance (CR) absorption. Phenomenologically, it would be the so-called photo-voltaic effect which always is the result of inner spatial asymmetry of an excited system [37].

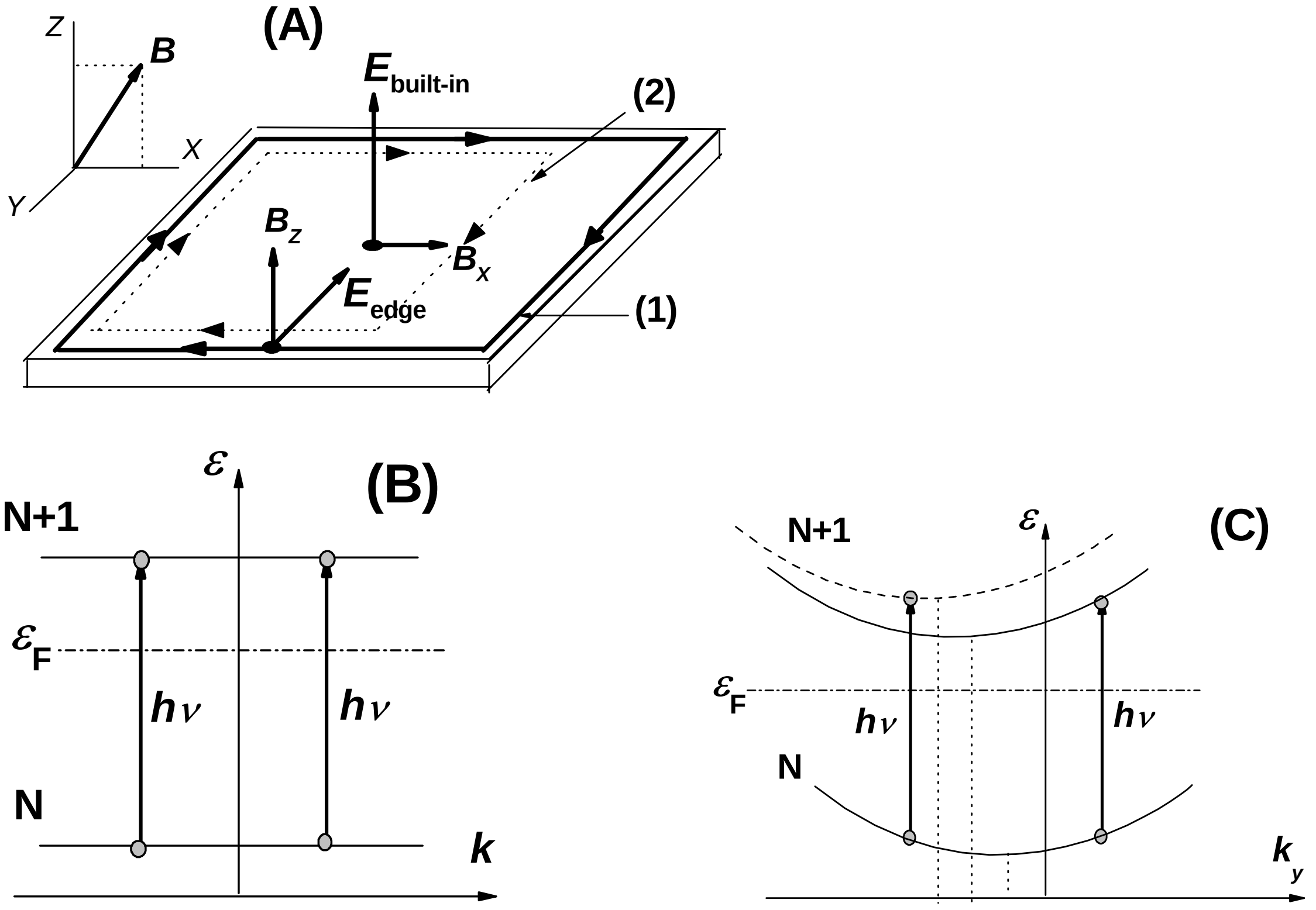

**Fig. 2. Schematic illustration of how the long-range orbit-like quantum states could emerge in asymmetric IQH system and of how they could be detected. (A) The scheme of in-plane spontaneous one-dimensional currents in unbiased IQH system with asymmetric confining potential in presence of tilted quantizing magnetic field: (1) true Laughlin-Halperin's edging currents, (2) Laughlin-Halperin-type currents in the system interior. (B) CR transitions in conventional IQH system (C) CR transitions when the shift of Landau subbands is a function of system toroidal moment as well as of the Landau quantum number.**

However, to be sure that detected responses (if any) are truly related to the Laughlin-Halperin-type states, we need in a rigorous criterion that would allow us to distinguish them from any other ones. To find the criterion, let us focus on the following remarkable thing. If the electrons' *X*-coordinates truly correlates with their wave vectors in accordance with the above-mentioned relation, then the electrons' velocities ($v_y(k_y)$) appear spatially-ordered in the *X*-direction. Moreover, the ordering is of such kind that it does not imply any spatial periodicity because $v_y$ is not a periodic function of $k_y$. If such ordering does exist, we should observe quite different local responses at different *X*-coordinates because, as is seen from Fig. 2C, the net current should be a function of $k_y$ and hence of $x_0$. That is just the criterion we need. Coming back to Fig. 2C, it is seen that CR conditions for electrons with positive velocities are fulfilled at lower magnetic fields than those for electrons with negative velocity. This means CR position as well as CR lineshape may also be sensitive to $x_0$.

## 4. Experimental details

In our experiments, we use an optically-pumped terahertz ammonia laser operated in a single-pulse regime. The energy of light quanta is 13.7meV, pulse duration is about 50ns and the incident intensity is about 200W/cm$^2$ [38]. Spatially-uniform unpolarized radiation is normal to the sample surface and does not thus induce an additional in-plane asymmetry. To provide a nonzero toroidal moment, the external magnetic field (up to 6.5T) is tilted from the normal by about $15^\circ$. The sample temperature is 1.9K, i.e. much less than the energy of light quanta. The high-speed in-plane responses are detected in a short-circuit regime with a $50\Omega$ load resistor.

Our IQH system is based on the MBE-grown semimetallic single quantum well structures of type InAs-GaSb, for which CR is expected at the magnetic fields of about 4.8T. In this type of quantum structures, the valance band of GaSb overlaps the conduction band of InAs by about 100 meV. To avoid a hybridization of these bands, a 15-nm-wide conducting layer of InAs is sandwiched between two 10-nm-wide AlSb barriers. Typical structure consists thus of a thick GaSb buffer layer followed by this sandwich and capped by a 20-nm-wide GaSb protecting layer. Under these conditions, Fermi level is well above the first quantum-size level but below the second level. Low-temperature electron sheet density is as high as $1.4 \cdot 10^{12}\,\mathrm{cm}^{-2}$ with the mobility of about $10^5\,\mathrm{cm}^2/\mathrm{Vs}$.

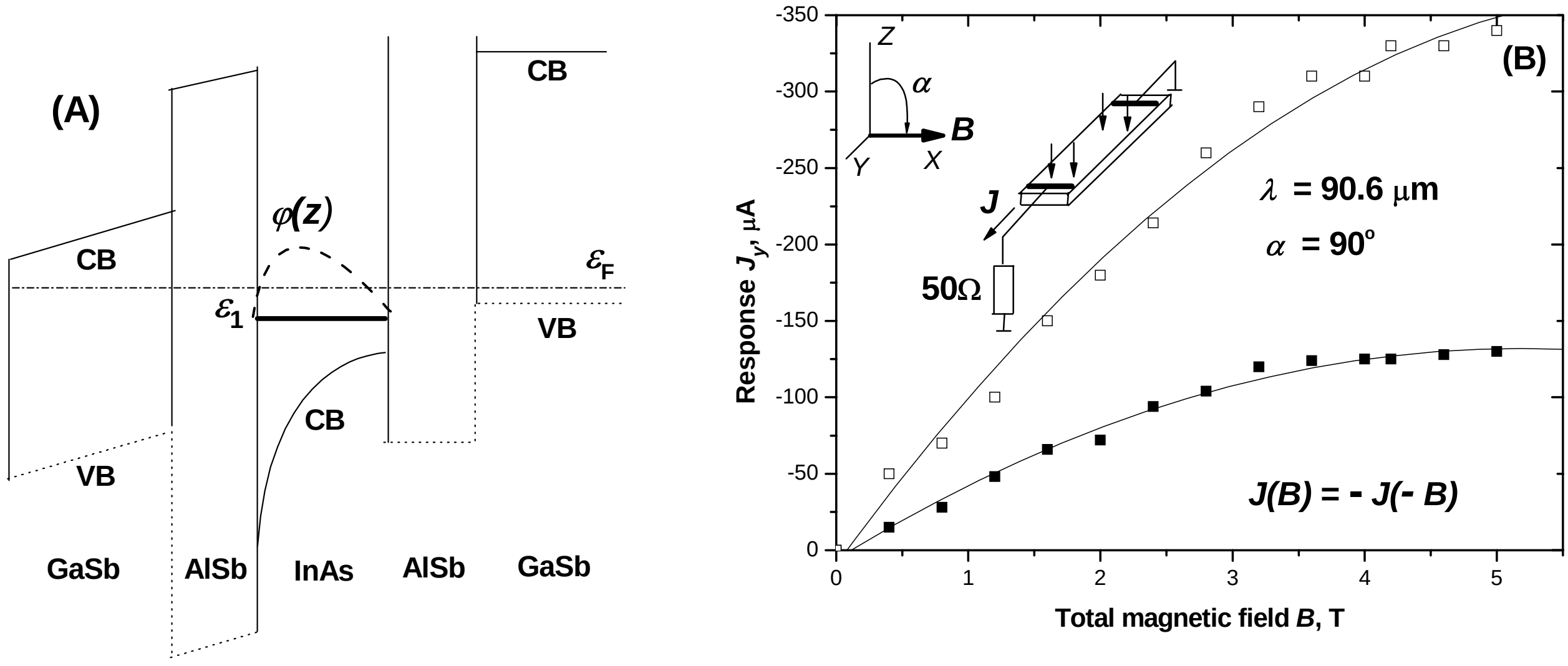

**Fig. 3. Evidence for a built-in electric field in the structures studied. (A) The energy-band diagram of semimetallic InAs-GaSb single quantum well structure with two AlSb barriers. Dotted line shows schematically the electron density shifted toward the more charged interface. (B) Test for the presence of built-in field. The inset shows experimental geometry. The outcome is shown for two samples with different growth parameters. Solid lines are a guide for the eye.**

Schematically, the energy band diagram of our system is shown in Fig. 3A where the asymmetry of confining potential is supposed to be due to the penetration of surface potential into the well [39]. The potential profile across the well is rather exponential than linear because of the high electron density. As a result, the built-in electric field is a sharp function of *Z*-coordinate.

To be sure of the presence of built-in field, all structures were preliminary tested through the method shown in the inset of Fig. 3B. Actually, it is the photo-voltaic measurements in a classic regime when the external magnetic field is non-quantizing. The test is based on the simple and reliable idea that the in-plane magnetic field alone, as a pseudo-vector, cannot provide an in-plane response which should be a polar vector. However, the cross product of in-plane magnetic field and built-in electric field is indeed a polar vector, that is, the system toroidal moment. Therefore, an in-plane response could occur, at least for symmetry reasons. In the test, we use the samples of $5 \times 12\,\mathrm{mm}^2$ with a single pair of ohmic contacts. Typical outcome is shown in the main graph of Fig. 3B. It is seen that an in-plane response does occur. As expected, it increases with increasing of magnetic field and is sensitive to the MBE-growth parameters those could alter the built-in field.

## 5. Test for the presence of macroscopic-scale quantum orbits

To test the presence of Laughlin-Halperin-type states in accordance with the selected criterion, we take a relatively large sample ($19 \times 12\,\mathrm{mm}^2$) with the three short contact pairs (1mm in length) belonged to different sample domains and remote from each other at a distance of no less than 7mm (Fig. 4A). To avoid a direct influence of sample edges, each contact is remote from the closest edge by about 1mm and for the sake of convenience all pairs are numbered from left to the right. Fig. 4B shows CR spectra obtained. It is clearly seen that despite exactly the same experimental conditions, they indeed differ drastically from each other so that they have even the opposite signs at the opposite sample ends. To be sure that we are truly dealing with a spatial ordering, we reverse magnetic field and repeat our measurements. The responses appear spatially-redistributed roughly in accordance with the following relations: $J_1(-B) = -J_3(B)$; $J_2(-B) = -J_2(B)$; $J_3(-B) = -J_1(B)$. That is just what should be in the case of the spatial ordering because reverse of magnetic field should result in a reversed shift of Landau subband in *k*-space (see Fig. 2C).

It should be noted that, by definition, we have realized a peculiar macroscopic quantum phase in which translational symmetry is lacked insofar as the local responses are not spatially periodic in the *X*-direction [40]. The phase can thus be regarded as the result of a continuous quantum phase transition from the quantum Hall state of matter, which is governed by the system toroidal moment. The characteristic feature of the phase is that it has the lowest symmetry among all quantum phases known up to now. In a sense, any system in the phase is reminiscent rather a gigantic single atom and this is just what we need to carry out the main test.

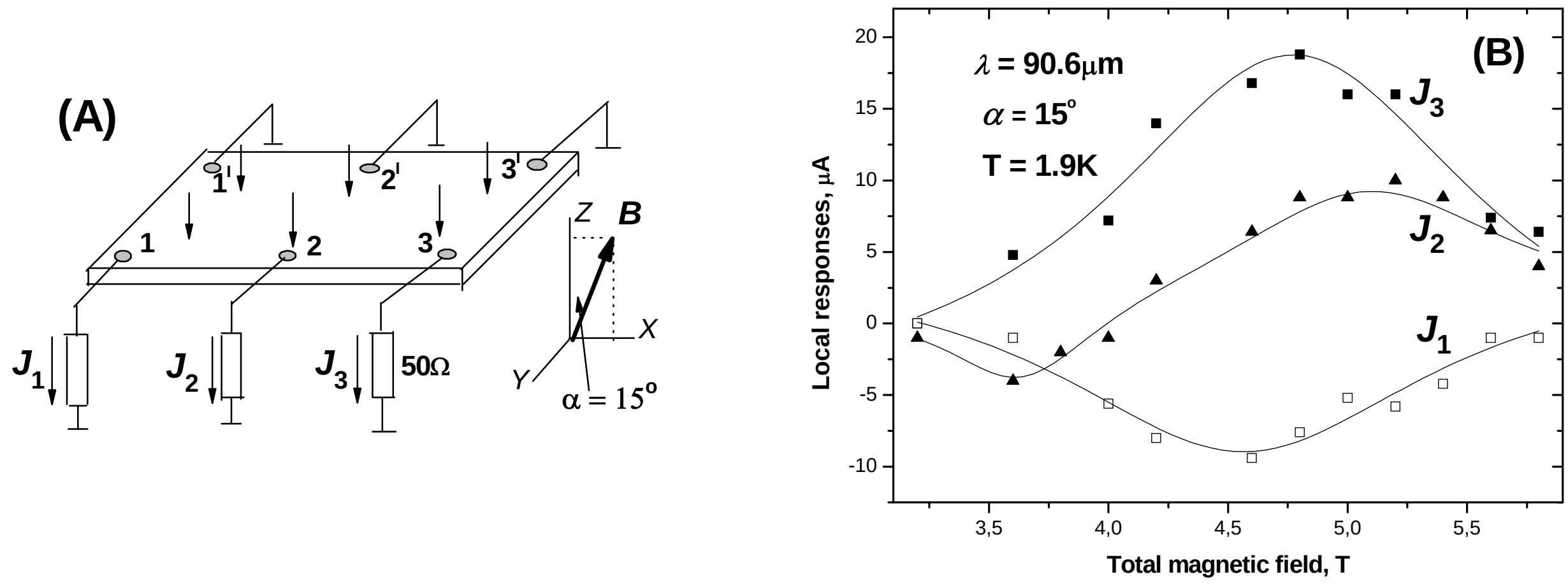

**Fig. 4. Evidence for Laughlin-Halperin-type states in asymmetric IQH system. (A) The scheme to detect local photo-responses from different sample domains. All contacts are identical and each one is 1mm in length. The distance between adjacent pairs is 7mm. The total sample length is 19mm. (B) CR spectra obtained from these pairs.**

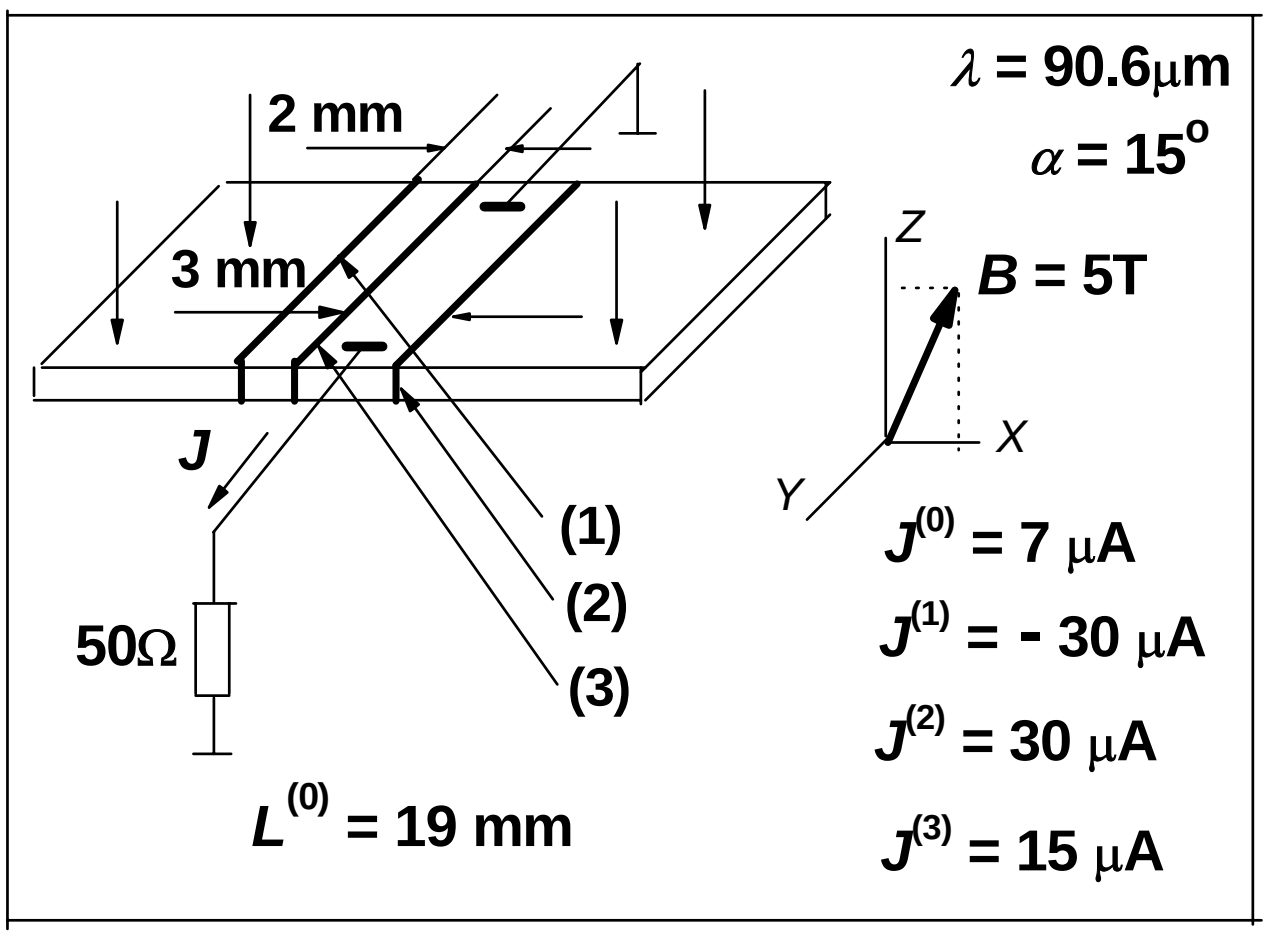

**Fig. 5. Evidence for high sensitivity of local responses to the position of remote sample edges. The responses from single short contact pair are measured after sequenced mechanical cuttings. The cutting lines are parallel to the *Y*-direction. Long arrows show their position with respect to the contact pair.**

However, the lack of translational symmetry automatically implies that any local response should be extremely sensitive to the position of system edges regardless of their remoteness in the *X*-direction. Thus, the presence (or absence) of this effect allows one to test once more our guess. To perform the test, we take a large sample ($19\times 12\,\mathrm{mm}^2$) but now with a single short contact pair (1mm long) centred in the *X*-direction (Fig. 5). At fixed tilted magnetic field ($B\approx 5\,\mathrm{T}$), we measure photo-responses each time the sample has become shorter in the *X*-direction due to mechanical cuttings. Three sequenced cuttings have been made in such a way that each new edge is remote from the contacts at a quite macroscopic distance of no less than 1mm (see Fig. 5).

The outcome is presented in the right-hand corner of Fig. 5 where the upper index denotes the number of cuttings before a given measurement. It is clearly seen that each cutting does change drastically the response so that even the sign may become reversed. This fact cannot be interpreted in terms of any trivial effects depended on the sample length and therefore we are truly dealing with a phase in which the Laughlin-Halperin-type states are responsible for the breaking of translational symmetry.

## 6. The main test

The apparent success of the test for the presence of macroscopic quantum orbits gives us the chance to realize the main test in accordance with the scheme in Fig. 1. Indeed, the role of *A*→*C* transitions could play CR optical transitions while the role of *C*→*B* transitions could play the intra-subband scattering-induced transitions. Here the characteristic time $\tau_{C \to B}$ is the so-called quantum relaxation time which is of the order of 0.3ps in typical structures and is always much shorter than the electron lifetime in higher Landau levels [41-42].

The strategy of the main test is as follows. If the nonlocal quantum transport does exist, then the efficiency of Bob's detection should not depend on whether one excites his own domain or the Alice's domain. However, in terms of everyday intuition this effect seems even a bit crazy because no reasons for any transport between these domains. So, the test consists of two experiments shown in Figs. 6A and 6B, respectively. In both cases, we use large samples ($19 \times 12\,\mathrm{mm}^2$) with a single short contact pair (1mm long) remote from the closest sample edge by about 1mm. Only one third of the sample is exposed to light. However, in the experiment 6A the contacts are inside the laser spot while, in the experiment 6B, they are remote from the laser spot at a distance as long as about 1cm.

Fig. 6C shows CR spectra obtained in both experiments and the result exceeds even the most optimistic expectations. The responses are indeed roughly the same despite the fact that any electric connection is impossible at least because the resistance of unexposed domain is as high as about 10kOhm, i.e. two orders higher than the load resistance. This means the nonlocal quantum transport does exist. Moreover we can roughly estimate the "speed" of the transport bearing in mind that this "speed" has nothing to do with a movement throughout the sample. Under the high laser intensity, it seems a good approximation that the characteristic time necessary to "overcome" the distance of 1cm is just the quantum relaxation time and we thus obtain the "speed" of the order of $3\times10^{12}$ cm/s, i.e. two orders higher than the speed of light.

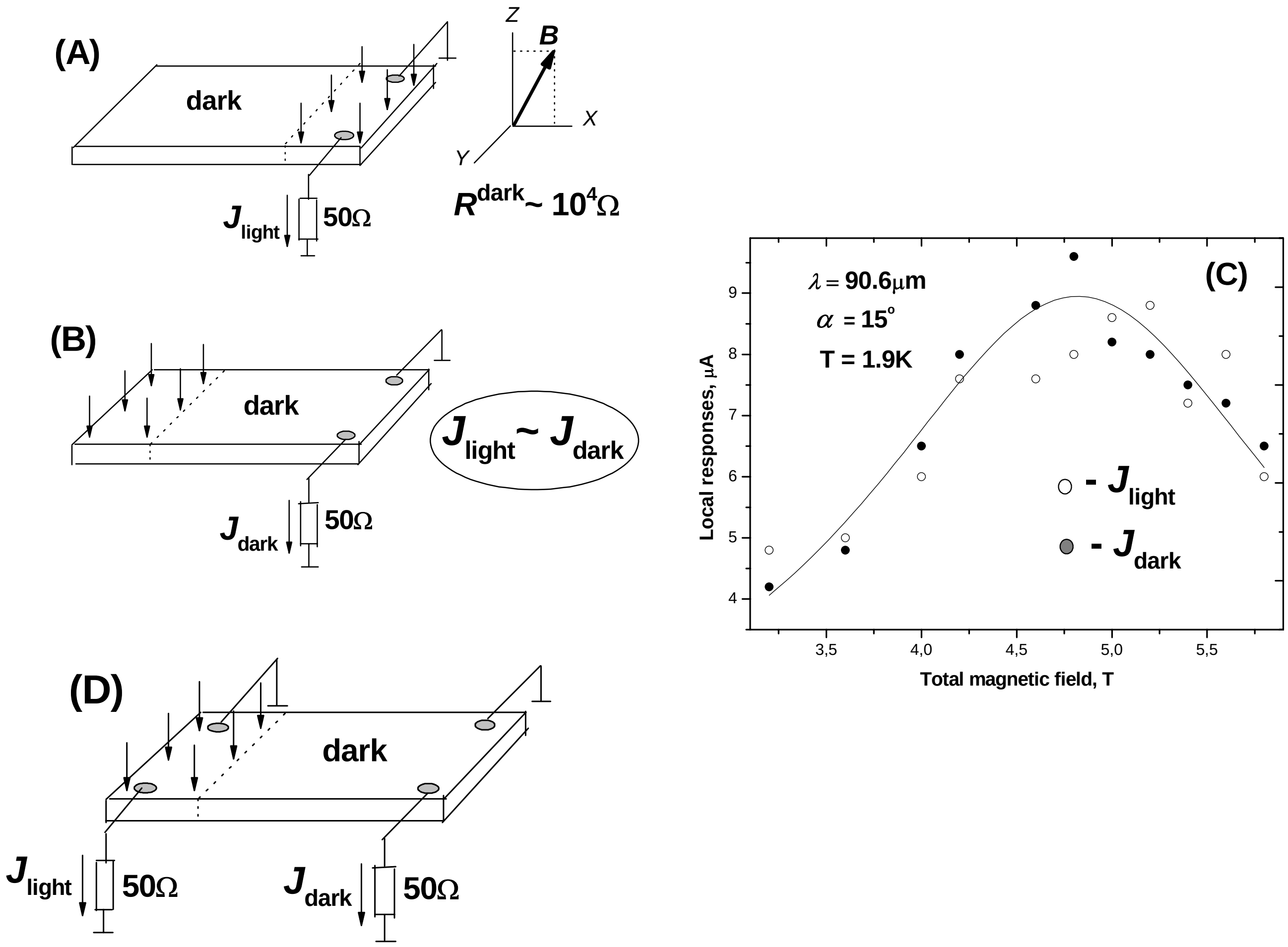

**Fig. 6. The experimental test of nonlocal single-particle transport. (A) A one third of the sample is exposed to light while the contact pair is inside the laser spot. (B) The same experiment but now the contact pair is remote from the laser spot at a distance of 1cm. (C) CR spectra obtained in the experiments A and B. (D) The scheme of synchronous detection of the responses from both the illuminated pair and the pair in the dark.**

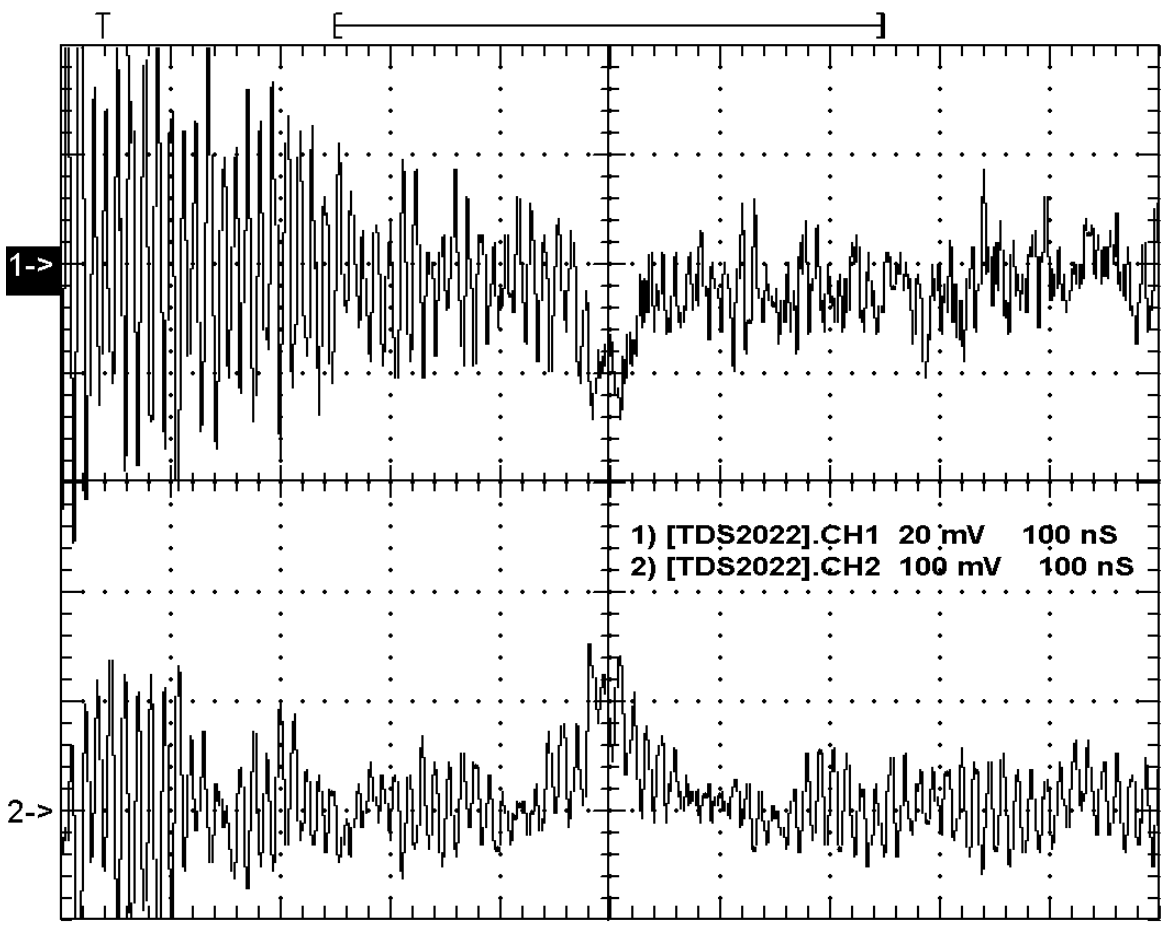

**Fig. 7. The outcome of the synchronous detection. Upper track – the signal from illuminated pair (20mV/div.); lower track – the signal from pair in the dark (100mV/div.). Timescale is 100ns/div.**

To avoid any misinterpretation, we perform an additional experiment in which we use two contact pairs: inside and outside the laser spot (Fig. 6D). Fig. 7 shows typical tracks under synchronous detection of both responses. It is clearly seen that the *dark* response is even in a factor of five *higher* than the response from spotlit domain. This fact rules out the interpretation of the former in terms of a "secondary" effect with respect to the latter. AS expected, under reversed magnetic field the responses are redistributed as if the whole sample is still exposed to light: $J_{dark} \approx -J_{light}$; $J_{light} \approx -J_{dark}$. As it is also seen from the oscillogram, there is no delay between the responses with an accuracy of about 20ns. It follows that, to have a little chance to be relevant, any transport from laser spot must be of a ballistic character with a speed of no less than $10^8$ cm/s and with a mean free path of *more than 1cm*. This is absolutely impossible because of a gigantic number of various scatterers.

Finally, we repeat both experiments (6A and 6B) but in the classic regime when the magnetic field has only *X*-component. No responses are observed in the experiment 6B despite the fact that the responses in the experiment 6A are as high as about 0.5mA, i.e. two orders higher than in quantum regime.

## 7. Fundamental consequences

Although we have observed a nonlocal quantum transport at a relatively short distance, the very such fact has truly fundamental consequences. First of all, this observation compromises the mysticism in quantum mechanics insofar as the requirement of Lorentz invariance is inapplicable to any version of quantum theory and the quantum dilemma is thereby a fiction. It follows immediately that the mysticism of orthodox quantum theory no longer has advances against the realism which currently is embodied in de Broglie-Bohm pilot-wave theory. Moreover, now the situation is just opposite. The realistic quantum theory has uncontested advances against the mysticism because it opens the door for the Einstein's dream of a unified realistic foundation of all physical theories and allows one to avoid quite naturally the problem of "switching" from classic realism to quantum mysticism.

Surely our observations could potentially be interpreted in terms of orthodox quantum theory. Indeed, to be understandable, the description of our *gedanken* experiment we have carried out just in these terms. For well-known historical reasons, this way of description is certainly more comfortable for us and, in a purely utilitarian aspect, it could be fruitful. However, in a fundamental aspect, further escalation of mysticism seems unacceptable because quantum mechanics remains thus a kind of outcast in the family of natural sciences despite its unprecedented success in prediction of various effects.

At the same time, the new status of quantum mechanics as a realistic theory allows it to claim something more than just the prediction of experimental observations. The point is what we actually observe in our experiments is a *new dynamics* that allows a spatially-discontinuous movement beyond the relativity. However, throughout the whole history of physics the current concept of spacetime is always determined by dynamical theory which currently is the all-embracing one in the sense that any other dynamical theories can be regarded as limiting cases. For example, at the beginning of twentieth century the Newton's dynamics was replaced by relativistic dynamics. As a result, the concept of linear Cartesian space as well as the concept of absolute time was replaced by the new concept of curvilinear Cartesian space together with the concept of relativity of time. However, the concept of strict continuity as well as the concept of strict locality was maintained. Accordingly, the paradigm of continuous movement of a dimensionless material point was also maintained. Now we face a new dynamics that can be described by a quite realistic theory and therefore a new "quantum" concept of spacetime is required instead of the current relativistic concept.

As for the quantum concept of time, the situation seems quite evident. *De-facto*, the spatially-discontinuous movement may well be regarded as the procedure of instantaneous synchronization of remote clocks. Moreover, this procedure allows one to avoid any causality paradoxes, such as Tolman's paradox [43-44], because all these paradoxes rest on the Lorentz transformations which are irrelevant to this kind of movement. Surprisingly, we thus come back to the Newton's concept of absolute time, which certainly is extremely comfortable insofar as it is adequate to our intuitive perceptions.

However, as for the quantum concept of space, it seems quite natural to expect that it should be much more counter-intuitive taking into account the counter-intuitive character of quantum phenomena at least from the viewpoint of our everyday intuition. And this is indeed the case. The point is that this concept already exists. It is the Bohm's concept of implicate order. At present, this concept is generally regarded rather as a mind game and it is hardly surprising the very de Broglie-Bohm interpretation of quantum mechanics is regarded to be strongly compromised by the above-mentioned quantum dilemma. However, in the view of the disavowal of quantum dilemma, the concept of implicate order deserves much more attention. Originally, the basic motive for this concept is the fact that the Cartesian notion of spatial order violates an essential content of quantum mechanics, that is, the discontinuity and the nonlocality though, to be exact, Bohm always contemplated the non-causal EPR-like nonlocality.

In essence, the concept of implicate order essentially does not treat a quantum particle as being separated from the other distant particles. Rather, they all are a single system which has an objective quality of unbroken wholeness. This means each part of the system contains an implicit order essentially similar to that of the whole system just like each part of a hologram contains implicit information about the whole image and, under certain conditions, could reproduce it but may be not so distinctly. The notion of a dimensionless material point is violated and this contrasts with explicate order of Cartesian space where each thing lies only in its own particular region and outside the regions belonging to other things. To understand nonlocal phenomena, each particle should be regarded "…as a projection of a 'higher-dimensional' reality, rather than as a separate particle, existing together with all the others in a common three-dimensional space…" and the notion of "higher-dimensional reality" should be in a close connection with familiar quantum-mechanical notion of multi-dimensional configuration space.

Thus, relativistic notion of spacetime appears a limiting case of a deeper quantum notion just like the classic notion of spacetime linear is a limiting case of the relativistic notion. If this hierarchy will be generally recognized, then quantum mechanics transforms from a merely mathematical tool to predict experimental observations into the most fundamental of current physical theories. Otherwise, we will face the paradox of an active influence on the past and moreover we will face further expansion of so-called quantum mysticism, now toward the macroscopic world.

**Acknowledgements**

The author would like to thank R. Chiao (University of California) for useful discussion. Gratefully acknowledged are B. Y. Meltser and S. V. Ivanov (Ioffe Institute) for supplying by the MBE samples. The also acknowledged are S. D. Ganichev (University of Regensburg) and Y. V. Terent'ev (Ioffe Institute) for invaluable contribution into the experimental equipment.